# Do methods matter in the meta-analysis of partial correlation coefficients?

T. D. Stanley[1], Petr Cala[2], Hristos Doucouliagos[3], Zuzana Irsova[4], and Tomas Havranek[5]

**Abstract**

Recent studies have demonstrated that conventional meta-analyses of partial correlation coefficients (PCC) are biased. Several adjustments have been shown in simulations to reduce these small-sample biases to negligibility. While many meta-analyses of partial correlation coefficients are conducted each year across several disciplines, the practical importance of these issues remains unknown. To address this question and to offer advice for applications, we survey 172 economic meta-analyses of PCCs. We find that small-sample biases are negligible in practice. However, some publication selection biases remain. Although Fisher's *z* transformations have often been recommended, they reduce neither small-sample nor publication selection biases relative to conventional random effects. Both the unrestricted weighted least squares (UWLS) and the Hunter-Schmidt (HS) estimators produce smaller, arguably less biased, estimates of the mean PCC in these applications than either random effects with or without Fisher's *z* transformations. These findings offer practical guidance for any discipline that meta-analyzes partial correlations.

*Web appendix: https://meta-analysis.cz/pcc_survey/*

[1] Department of Economics, Deakin University, 221 Burwood Highway, Burwood, 3125, Victoria, Australia and Meta-Research Innovation Center at Stanford. Email: stanley@hendrix.edu.

[2] Institute of Economic Studies, Faculty of Social Sciences, Charles University, Prague. Email: cala.p@seznam.cz.

[3] Department of Economics, Deakin University, 221 Burwood Highway, Burwood, 3125, Victoria, Australia. Email: douc@deakin.edu.au.

[4] Institute of Economic Studies, Faculty of Social Sciences, Charles University, Prague, and Meta-Research Innovation Center at Stanford. Email: zuzana.irsova@fsv.cuni.cz. Irsova acknowledges support from the Czech Science Foundation (grant no. 23-05227M).

[5] Institute of Economic Studies, Faculty of Social Sciences, Charles University, Prague, Centre for Economic Policy Research, London, and Meta-Research Innovation Center at Stanford. Email: tomas.havranek@fsv.cuni.cz. Havranek acknowledges support from the Czech Science Foundation (grant no. 24-11583S).

## Highlights

### What is already known?

- Conventional meta-analyses of partial correlation coefficients (PCCs) are biased, although the biases are relatively small in most cases.
- Several adjustments have been proposed to reduce these small-sample biases to negligibility.
- Many meta-analyses of PCCs are conducted each year.

### What is new?

- We survey the performance of alternative methods across 172 meta-analyses of PCCs.
- In practice, small-sample biases are trivial; however, likely publication selection bias remains.
- Although Fisher's *z* version of random effects is often recommended, it does not reduce estimated mean effects relative to conventional random-effects models.
- Both unrestricted weighted least squares (UWLS) and Hunter and Schmidt's approaches to the meta-analysis of PCCs produce more conservative estimates that are thereby likely to be less affected by widely known upward biases than random effects in applications.

### Potential impact for *Research Synthesis Methods* readers outside the authors' field

- Our comparative findings are likely to apply to all disciplines where one wishes to conduct a systematic review or meta-analysis of PCCs. Although our data come from economics, the key findings are relevant wherever PCCs are meta-analyzed with heterogeneous sample sizes, including health sciences, education, and psychology.

## 1. INTRODUCTION

Meta-analysis is used across the disciplines to integrate, summarize, and understand different research findings for a given area of research, hypothesis, or treatment. Meta-analyses are statistical analyses of statistical findings that are regarded to be at the top of the "pyramid of evidence".[1,2] However, they require a fully comparable metric (an 'effect size') that quantifies the sometimes very different research findings across numerous studies of the same phenomenon.[3-5] In economics and other social sciences, it is common to find that different studies of the same conceptual effect employ different outcome measures, treatment metrics, dependent and/or independent variable measures of the phenomenon in question. To circumvent this comparability issue, partial correlation coefficients (PCCs) are frequently used in the meta-analysis of economics research (e.g.,[6-8]). PCCs are also increasingly used in environmental science, management, and public health, making the choice of meta-analysis methods relevant well beyond economics. As a unitless measure of statistical effect (a correlation), PCCs can be sensibly compared across different measures of the dependent variable (for example, per capita GDP, GDP, and GDP growth) and different metrics of the independent variable (e.g., broadband access, cell phone contracts, or IT expenditures when technology-growth is being studied).

Recent studies cast doubt on the proper use of partial correlations in meta-analyses. Van Aert and Goos[9] claimed that using the 'wrong' formula for PCC's variance, $\left(1-r_p^2\right)\big/_{df}$, biases meta-analysis estimates of the mean effect. While Stanley and Doucouliagos[10] showed that conventional meta-analyses of PCCs (e.g. fixed and random effects) are biased, using this 'wrong' formula systematically and predictably reduces these biases. Usefully for practice, these biases tend to be small and become negligible with large primary study sample sizes. Stanley, Doucouliagos, and Havranek[11] find that both well-known and novel approaches to correcting publication selection bias (PSB) can reduce these small-sample biases to scientific negligibility.

These small-sample biases are caused by a mechanical correlation of PCC estimates with their variances, which, in turn, affects any inverse-variance weighted average. Traditionally, the solution to a correlation between effect sizes and their standard errors (SE) is to employ the Fisher's $z$ transformation to random effects. Although simulations confirm that Fisher's $z$ transformation can do much to reduce this bias, Stanley, Doucouliagos and Havranek[11] find that a simple adjustment to degrees of freedom allows the unrestricted weighted least squares (UWLS) to reduce these biases further. Stanley, Doucouliagos and Havranek[12] have also shown that the well-known but infrequently employed meta-analysis method developed by Hunter and Schmidt[13] also works well to practically eliminate these same small-sample biases seen in the meta-analysis of bivariate correlations. Although these biases are well-established in theory and simulations, will they matter in practice? This is the central question addressed by this paper.

Empirical confirmation is important because no simulation can capture the full complexity of actual meta-analytic data, and simulation-based recommendations that fail in practice could mislead hundreds of applied meta-analyses each year. In particular, the typical area of research surveyed here uses a wide dispersion of sample sizes. The average of median sample sizes is 958 and the average interquartile range is 3,474, representing both large average sample sizes and wide dispersions among the sample sizes within the survey's typical PCC meta-analysis. This contrasts sharply with previous simulations where the sample size was held constant for each study in the meta-analysis, and yet the identified small-sample biases became trivial at sample sizes of 400 or larger.[10] In addition to the wide distribution of sample sizes, meta-analyses of PCCs typically have high heterogeneity and are often afflicted with notable PSB. Even though heterogeneity and PSB have been simulated in Stanley, Doucouliagos and Havranek,[11,12] the complexity and nuance of typical PCC meta-analyses make it essential to assess these alternative methods in practice.

From a large set of meta-analyses of economics research, we have identified 172 meta-analyses that were conducted using PCCs as the effect sizes.[14,15] Below, we investigate this large set of empirical economic effects (63,730) to see whether these alternative meta-analysis methods cause notable differences to the interpretation of the findings for these areas of research. Because hundreds of meta-analyses are conducted each year using PCCs, guidance about which methods are best is needed, yet no study has systematically compared these estimators across a large body of empirical meta-analyses.

The purpose of this paper is to investigate whether alternative meta-analysis methods matter, practically or scientifically, in the synthesis of partial correlations. Typically, economics meta-analyses involve hundreds of estimates, many of which come from studies that used hundreds, thousands, or even millions of observations. With large sample sizes, these small-sample biases are expected to become scientifically negligible. However, with both small- and large-sample studies mixed together, the relative advantages or biases of alternative meta-analysis methods are unknown. We fill this gap in our understanding by investigating 172 economic meta-analyses containing 63,730 estimates and their standard errors (SE) to see if the use of alternative meta-analysis methods has any notable effects on the summary estimates of PCCs. In the process, we find that the typical effect in this literature is very small. The median PCC is approximately 0.06, which in many contexts would fall below the smallest effect size of interest.

## 2. META-ANALYSIS OF PARTIAL CORRELATIONS

Partial correlations are often employed to solve a common problem found in social science research. Different studies of the same economic phenomenon will frequently quantify the dependent variable and/or the independent variable using fundamentally different units of measurement. For example, per capita GDP, productivity increases, and GDP growth rates are all used by empirical studies when studying the effect that the further adoption of information and communication technology (ICT) has on a country's 'economic growth'.[7] Substantively different scales, metrics, or measures cannot be statistically or meaningfully combined. However, this "apples and oranges" dilemma conflicts with systematic reviews' imperative to include all relevant research findings on a given issue or phenomenon.[4,5,16,17]

Partial correlations offer a straightforward escape from this dilemma. Any estimated regression coefficient can be converted to a partial correlation coefficient, $r_p$, by:

$$r_p = t \Big/ \sqrt{t^2 + df}\,, \tag{1}$$

Gustafson.[18] $t$ is the conventional $t$-test for the statistical significance of the effect in question (e.g., ICT growth) in the explanation of the dependent variable (e.g., GDP growth), and $df$ is the degrees of freedom available to the estimated multiple regression. $r_p$ measures the statistical strength and direction of a linear relationship between the dependent phenomenon (e.g., GDP growth) and the variable of interest (e.g., AI investment), after factoring out the effects of other independent variables (e.g. trade and education). $r_p^2$ is the proportion of the variation in the dependent variable attributable to variation in the target independent variable after accounting for the effects of all other independent variables.

Eq. (1) provides comparable effect sizes as long as the different dependent variable metrics (or independent variable metrics) are all considered to measure the same general economic phenomenon (e.g., economic growth). However, to conduct a meta-analysis, we still need the standard errors (SE) of these PCCs. Stanley and Doucouliagos[5] advocated the use of the same SE that is used to test whether the population partial correlation is zero ($H_0$: $r = 0$) because it is this test of statistical significance that will potentially be gamed for those who seek to report statistically significant findings (i.e., publication selection bias or PSB):

$$S_1 = \sqrt{(1 - r_p^2) / df} \tag{2}$$

Stanley and Doucouliagos.[5(p25)]

Recently, Van Aert and Goos[9] criticized the recommendation of using Eq. (2) to calculate PCC's SE as the asymptotic variance of PCC's sampling distribution is widely known to be the square of:

$$S_2 = \sqrt{(1 - r_p^2)^2 / df} \tag{3}$$

Olkin and Siotani.[19] Thus, $S_2$ is the 'correct' SE if we confine our interest to the sampling variation of a *single* PCC estimate. However, the conventional 'correct' SE formula, by more faithfully capturing sampling variation, ironically induces larger inverse-variance weighting distortions and resulting meta-analysis biases. Thus, the 'correct' SE formula is the 'wrong' formula to use in meta-analysis. Furthermore, because $S_2 < S_1$ for all $|r_p| \neq \{0 \text{ or } 1\}$, $S_2$ can make some studies with statistically insignificant effects appear statistically significant. This is especially a problem in economics where there is often a jump in the frequencies of reported statistical tests just above the .05 threshold.[20,21]

Nonetheless, Van Aert and Goos[9] claim that using any formula other than Eq. (3) for PCC's SE "biases the results of meta-analysis." Stanley and Doucouliagos[10] show, to the contrary, that using $S_2$ in conventional meta-analysis (fixed effect, random effects, and unrestricted weighted least squares, UWLS) doubles the bias relative to $S_1$. Thus, for use in meta-analysis, $S_2$ is the 'wrong' estimate of the standard error as $S_2$ consistently produces larger biases. However, these biases are small-sample biases that exaggerate estimated mean effect sizes but decrease with larger sample sizes. Below, we investigate whether these different standard error/variance formulas make a notable difference in *practice*.

Conventional meta-analysis estimates the overall mean effect for a given area of research by inverse variance weighted averages, called fixed and random effects. The fixed-effect estimate (FE) of *r* is a weighted average of all the PCCs in an area of research with weights, $1/_{SE_i^2}$, where $SE_i$ can represent either $S_1$ or $S_2$. Fixed effects assume that there is a single common effect with no variation in 'true' effects. In economics, research results are known to be conditional upon different economic systems, institutions, laws, histories, customs, etc., creating systematic heterogeneity that is almost always identified by multiple meta-regression. Thus, FE should be rarely and cautiously used to summarize economics research.[i] Furthermore, observed heterogeneity is typically several times larger than what can be accounted for by SEs alone.[23]

Random effects (RE) explicitly accommodate heterogeneity using weights, $(1/_{(SE_i^2 + \hat{\tau}^2)})$, that include an estimate of the heterogeneity variance, $\hat{\tau}^2$. Because $\hat{\tau}^2$ is constant across all of the estimates within a meta-analysis, random-effects weights are moderated from the upweighting of the more powerful and more precise, $1/_{SE_i}$, estimates. However, many studies have shown that RE is more biased in the presence of PSB than FE and the unrestricted weighted least squares (discussed below), and there is much evidence that publication bias is often substantial in economics.[5,14,15,20,23-32]

Like FE, the unrestricted weighted least squares (UWLS) uses inverse variance weights $(1/SE_i^2)$. Unlike FE, UWLS does not assume a common effect and estimates a proportional variance, γ, which mechanically and automatically adjusts for any excess heterogeneity when present. UWLS is estimated by $\alpha_1$ in the simple meta-regression model:

$$t_i = \frac{r_{p_i}}{SE_i} = \alpha_1 \left(\frac{1}{SE_i}\right) + u_i \qquad i = 1, 2, \ldots, k \qquad (4)$$

Where $k$ is the number of PCCs combined into the meta-analysis. Any standard regression software automatically reports UWLS, $\hat{\alpha}_1$ , its standard error, *t*-test, and confidence intervals when: *t*-values, $r_{p_i}/SE_i$ , are specified as the dependent variable, precision, $1/SE_i$ , is the only independent variable, and there is no intercept. Although UWLS will have the same point estimate as FE, the difference is that UWLS does *not* assume a common effect and will automatically accommodate observed between-study multiplicative heterogeneity, when present, in its estimated variance, $\hat{\gamma}/\Sigma\, ^1/_{SE_i^2}$; where $\hat{\gamma}$ is the MSE of meta-regression Eq. 4. UWLS is a variety of random effects where the estimates' variances are allowed to be proportional to $SE_i^2$.

Some have associated UWLS with the Knapp-Hartung (KH) method of moderating RE's heterogeneity variance estimate's ($\hat{\tau}^2$) well-known bias and high uncertainty.[33,34] KH uses the traditional RE additive inverse variance weights, $^1/_{(SE_i^2 + \hat{\tau}^2)}$, estimates the mean exactly as does RE, and lastly inflates RE's SE to compensate for $\hat{\tau}^2$'s downward bias using the square root of the weighted mean squared errors calculated from RE's estimate of the mean and RE's weights. Although KH and UWLS share the general WLS approach to estimating excess variation, UWLS and KH are otherwise different in every practical way. UWLS and KH have different estimates of the mean effect with different uncertainties reflected in their different SEs, CIs, and test statistics, with the lone trivial exception that FE = RE = UWLS = KH when there is no heterogeneity (i.e., $\gamma = 1$ & $\tau^2 = 0$). HK's overdispersion estimate will also be very different than UWLS' γ. The association of UWLS with KH might have arisen from van Aert and Jackson[35] WLS justification for HK's inflation of RE's SEs. In contrast to HK's *ad hoc* adjustment to RE, UWLS is a fully formed classical regression model when variances are assumed to be proportional to $SE_i^2$ and thereby gains all of least squares desirable statistical properties. Without requiring normality, these desirable properties include the independence of UWLS's estimates of the mean and heterogeneity variance as well as BLUE (best,' or smallest variance, linear, and unbiased estimator) by the Gauss-Markov theorem.[36-38]

All three of these meta-analysis estimators of mean effect (RE, FE and UWLS) have small-sample biases, regardless of which formula for SE is used.[10,11] The reason for these biases is that the weights themselves depend on the estimated value of PCC, $r_p$. As discussed above, all conventional meta-analyses employ inverse-variance weights. Because PCC variances are proportional to $(1-r_p^2)$, or its square, recall Eqs. (2) and (3), positive sampling errors will receive larger weight, thereby biasing the meta-analysis estimate upwards.[10,11]

An obvious remaining question is: how can these biases be eliminated or at least reduced to scientific negligibility? Stanley, Doucouliagos and Havranek[11,12] offer several approaches to reduce these biases to scientific triviality. However, it should be stressed that these are *small-sample* biases; thus, they are virtually zero (i.e., < .01) without any adjustment whenever $n_j \geq 200$. Because many social science regressions are estimated from more than 200 observations, often very much more, it is an empirical question whether these biases should concern meta-analysts in these fields at all. Answering this question is a central purpose of this study. To do so, we must also consider alternative adjustments that reduce these small sample biases to scientific triviality. Lastly, it is important to note that these small-sample and publication selection biases both tend to increase the magnitude of the reported PCCs, trends that have been verified in simulations and by our below survey results. Thus, smaller observed mean estimates, averaged across 172 meta-analyses and 63,730 estimated PCCs, are consistent with less overall bias.

## 3. ALTERNATIVE META-ANALYSIS ESTIMATORS OF PCCS

We investigate the four most promising adjustments for these small-sample biases of the meta-analysis of PCCs: the RE and the UWLS mean estimates of PCCs transformed to Fisher's *z*, *REz & UWLSz,* Hunter and Schmidt's[13] approach to the meta-analysis of correlations, *HS*, and lastly an adjustment to the degrees of freedom using UWLS, $UWLS_{+3}$.

### 3.1 Fisher's *z* Transformations: *REz & UWLSz*[ii]

The usual approach to avoid a mechanical correlation between estimated correlations and their variances is to first use the Fisher's *z* transform.[4] As Fisher[39] noted a century ago, what is true for correlations is true for partial correlations after degrees of freedom are appropriately adjusted. Fisher's *z*s are approximately normally distributed and their variance, $1/(n_i-3)$, does not depend on the specific value of Fisher's $z$. Thus, the cause of meta-analysis' small-sample bias is avoided as the variance of Fisher's zs is independent of the specific values of Fisher's $z$. In brief, the standard solution is to convert all PCC's to Fisher's $z$, $z = 0.5 \cdot \ln[(1 + r_p)/(1 - r_p)]$, calculate the random-effects (or UWLS) estimate of the mean, $\bar{z}$, and lastly convert this back to PCCs, $REz = \frac{exp(2\bar{z})-1}{exp(2\bar{z})+1}$

.

### 3.2 The Hunter-Schmidt Approach to the Meta-Analysis of Correlations

Hunter and Schmidt[13] offered an alternative meta-analysis approach (HS) to bivariate correlations, which they argued to be superior to RE$z$. HS uses the sample size, $n_i$, of each study as the weights. Thus, like Fisher's $z$, HS avoids any dependence arising from the weights being dependent on the estimated effect sizes and thereby their sampling errors. The HS meta-analysis estimate of the mean partial correlation is:

$$\overline{r_p} = \Sigma(n_i r_{p_i}) / \Sigma(n_i) \qquad i = 1, 2, \ldots, k. \qquad (5)$$

The SE of HS is not calculated in the conventional manner by the inverse of the sum of the weights, as RE and FE do, but rather as:

$$SE_{\overline{r_p}} = SD / \sqrt{k} \qquad (6)$$

Where PCC's standard deviation, $SD$, is computed by the square root of the weighted sum of squared deviations from the mean, $\overline{r_p}$:

$$SD^2 = \Sigma\left(n_i (r_{p_i} - \overline{r_p})^2\right) / \Sigma(n_i) \qquad i = 1, 2, \ldots, k \qquad (7)$$

[13,40].[iii] Because HS uses sample size rather than inverse variance weights, it further addresses the concern that inverse-variance weights can be distorted when reported standard errors reflect modeling choices rather than pure sampling uncertainty.[41]

### 3.3 Adjustment to Degrees of Freedom: *UWLS+3*

Recall by Eq. (1) that PCCs are calculated from the estimated regression coefficients' *t*-values and degrees of freedom, *df*. Because degrees of freedom are so central to the use of PCC in meta-analyses, perhaps a simple adjustment to the degrees of freedom will correct the small-sample bias of conventional meta-analyses. Simulations confirm that by merely adding 3 to *df* in Eq. (1) the small-sample meta-analysis bias is reduced to scientific negligibility.[11] We call the resulting weighted average '$UWLS_{+3}$.'

To be clear, $UWLS_{+3}$ substitutes degrees of freedom that are three larger than the multiple regression's degrees of freedom into PCC's transformation formula, Eq. (1), and uses $S_1^2$, Eq. (2), as the variance. That is, $UWLS_{+3}$ first calculates each PCC as:

$$r_p^{+3} = t \Big/ \sqrt{t^2 + df_{+3}} \tag{8}$$

for $df_{+3} = n - s + 2$ with $s$ as the number of independent variables in the multiple regression. Then, the meta-regression of $r_p^{+3}/S_1$ on $1/S_1$ is run, Eq. (4), to obtain the $UWLS_{+3}$ estimate. In their simulations, Stanley, Doucouliagos and Havranek[11] found that $UWLS_{+3}$ has smaller bias than REz and other conventional meta-analyses of PCCs.

### 3.4 Upward Bias Hypothesis: Comparing Alternative PCC Estimators

The known biases in the meta-analysis of partial correlation coefficients operate asymmetrically. That is, both small-sample bias and PSB systematically inflate meta-analysis estimated mean effect sizes. PSB is "one of the strongest findings across the sciences",[42(p370)] and dozens of surveys have found evidence of publication selection bias widely throughout the sciences, especially economics.[23,31] No analogous mechanism is known that would cause a typical meta-analytic estimator to *underestimate* the mean PCC systematically. In the absence of countervailing downward pressures, the direction of net bias is therefore one-sided: all estimators are plausibly too large, and the estimator that is, on balance, smallest is the one most consistent with the least upward distortion (*i.e.,* the *upward bias hypothesis).*

To clarify, bias is a theoretical property, equal to the difference between the expected value of an estimator's distribution and the value of the population parameter being estimated. Bias cannot be inferred from a single observed estimate nor does the fact that one estimator is smaller than another in a single application imply that it has smaller bias. To do so would be to confuse potential random sampling error or the effects of a singular realization of random heterogeneity as established fact. However, when an estimator is applied repeatedly across a large and diverse set of independent meta-analyses (172), its grand average converges to its theoretical expected value by the law of large numbers. That is, such grand averages are the empirical counterparts of expected values, which are equal to the 'true' mean effect plus any biases. An estimator that is consistently smaller, on average, across this breadth of research areas is therefore consistent with having a lower expected value, and, under the upward bias hypothesis, lower expected value implies less upward bias. This is precisely the kind of limited inference that aggregate empirical evidence licenses.

We make no claim that one can *infer* smaller bias by observing smaller average estimators. To do so would be to commit the fallacy of 'affirming the consequent.' One can never make an inference from empirical evidence to universal or theoretical 'truth.' Hume proved that this is impossible when discussing the problem of induction.[43,44] What empirical evidence can do is to provide corroborating support for a hypothesis when the direction of the evidence aligns with its predictions and when no plausible competing explanation accounts for the pattern as well. Furthermore, as we will see below, the empirical patterns seen in our survey are fully consistent with simulation studies with known small-sample and PSB biases.

## 4. AN ILLUSTRATION

Stanley *et al*.[7] conducted a meta-analysis of the partial correlations of the effect of information and communication technology (ICT) on economic growth. Although there is a strong consensus that the adoption and spread of these technologies are beneficial to the economy, some economists have questioned their practical effect. "You can see the computer age everywhere but in the productivity statistics".[45(p36)] Also, there is no universal measure of economic growth or of ICT. ICT consists of cell phones, broadband or other internet services, computer hard- and software, and now AI, while the most common economic outcomes are: GDP per capita, GDP growth rates, and increases in productivity. Economists who use productivity believe that increases in productivity will be reflected in economic growth, perhaps with a lag. Thus, with several different scales and ways to measure ICT and its economic impact, partial correlations are needed to measure the overall economic effect of ICT.

Using the data in Stanley et al.,[7] conventional random effects estimate the mean partial correlation between increases in ICT and economic growth as: 0.246, 95% CI (0.227, 0.265), and $k = 416$. Using REz does not notably affect these values: 0.241: 95% CI (0.230, 0.270), suggesting that RE has a very small, small-sample bias, 0.005, which is consistent with simulations.[10,11] According to conventional guidelines, both estimates are interpreted as 'small' effects ($.1 \leq r_p \leq .3$).[46] UWLS's estimate of the mean effect is 20% smaller, 0.195; 95% CI (0.177, 0.212) but remains what is widely considered a small effect. UWLS$_{+3}$ reduces UWLS by a trivial amount, 0.194; 95% CI (0.176, 0.211) but is 0.052 smaller than random effects. Lastly, both UWLSz and the Hunter and Schmidt (HS) approach find yet smaller estimates, 0.179; 95% CI (.162; .196) and 0.172; 95% CI (0.158, 0.187), respectively, which is 30% smaller than RE. If instead we were to use the variance and SE formula known to cause larger biases, $S_2$ from Eq. (3), the results are only slightly larger than those from $S_1$: RE=0.253; 95% CI (0.233, 0.273) and UWLS=0.198; 95% CI (0.184, 0.213). Also note that these ICT estimates line up in the exact order as seen in simulations where exact biases are known and align perfectly with the average magnitude of each estimator.[11,12] Although all these estimates are typically considered to be small, a 20-30% reduction in estimated effect size can have notable consequences in application for some moderators and meta-analyses, and it could, in some areas, alter policy recommendations and actions.

Although decreases of small-sample biases are responsible for some of these reductions (recall the very small difference between UWLS and UWLS$_{+3}$ and between RE and REz), reductions in publication selection bias can explain why UWLS and HS estimates are noticeably smaller than RE and REz. That is, the pattern and differences among these estimators' average magnitudes is consistent with the notion that PSB is the more severe bias in this area of research. This interpretation is also confirmed by tests for PSB. Both the Egger test and the proportion of statistical significance test (Egger: z= 5.97; p<.0001; PSST: z= 6.73; p<.0001) show clear evidence of PSB. The proportion of statistically significant test (PSST) is a test for the presence of PSB, and it cannot be interpreted as a small-study effect.[47] In general, UWLS has been shown to be less biased than RE when there is publication selection bias because its inverse variance weights are not moderated by RE's estimate of the heterogeneity variance.[24,25,27-32] As simulations have established, HS is expected to have smaller publication selection bias, because its weights are sample sizes, and PSB tends to be larger for small-sample studies.[12]

Of these alternative estimates of the mean partial correlation of ICT and growth, UWLSz and HS are the smaller, a pattern consistent with past simulations when there are both PSB and small-sample biases exaggerating these MA estimators. Nonetheless, it is important to recognize that none of these alternative methods produce differences sufficiently large to change how researchers interpret the overall, unconditional effect of ICT on economic growth.[iv] In all cases, we find that ICT has a small positive average effect on economic growth regardless of which meta-analysis estimator is used.

## 5. DATA

Our data were extracted from a large meta-research survey of statistical power and publication selection bias published in top-rated economics journals, as well as other scholarly journals, books, and unpublished working papers.[15] An extensive literature search for meta-analyses of economics research was conducted across: Econlit, Scopus, and Google Scholar. When the data were not publicly available (109 meta-analyses), 74% of the contacted authors shared their data.[15] This process resulted in 368 meta-analysis datasets with 167,753 estimated effect sizes of all types. From these meta-analyses, we identified 172 meta-analyses that synthesized partial correlations as their effect sizes. Together, these 172 meta-analyses contain 63,730 estimated PCCs and their SEs extracted from 6,191 papers (or approximately 371 estimated PCCs per meta-analysis with 10 per study). However, it is the sample sizes of the primary studies that drive the small-sample biases documented in previous papers. The median of the median sample sizes is 210 while the median of the mean sample sizes is over twice as large, 571, reflecting highly skewed distributions of sample sizes. This is further seen in the average interquartile range = 3,789-316= 3,473. Large sample sizes and wide dispersions among sample sizes both lessen small-sample bias.

Next, we investigate whether alternative approaches to the meta-analysis of PCCs make noticeable differences in application. Although these meta-analyses come from economics, the statistical properties of PCCs do not depend on the discipline; thus, the relative performance of estimators documented here should generalize to other fields where PCCs are used with similar sample sizes and heterogeneity.

## 6. RESULTS AND DISCUSSION

Table 1 reports the mean, median, standard deviation, frequency smallest, and mean squared error relative to PET-PEESE (the precision-effect test and precision-effect estimate with standard error) for these alternative meta-analysis estimates of the mean PCC across these 172 meta-analyses.[v] For ease of comparison, the median PCC in each area of research is forced to be positive. That is, for inverse correlations identified by having a majority of the PCCs < 0, we multiply all PCCs in that meta-analysis by -1. As a result, when present, both small-sample bias and PSB will be positive, exaggerating the magnitude of reported estimates and thereby the weighted averages of these selectively reported estimates. Thus, the smaller average estimates are consistent with smaller biases through the upward bias hypothesis.

Table 1: Descriptive summary statistics

| | Averages | | | | | | | | | Corrected |
|---|---|---|---|---|---|---|---|---|---|---|
| Statistic | Mean | RE1 | RE2 | UWLS1 | UWLS2 | $UWLS_{+3}$ | HS | REz | UWLSz | PP |
| Mean | 0.131 | 0.126 | 0.129 | 0.110 | 0.140 | 0.109 | 0.096 | 0.131 | 0.101 | 0.090 |
| Median | 0.097 | 0.087 | 0.090 | 0.061 | 0.069 | 0.061 | 0.056 | 0.088 | 0.056 | 0.040 |
| Std Dev | 0.115 | 0.116 | 0.117 | 0.128 | 0.175 | 0.127 | 0.109 | 0.123 | 0.117 | 0.171 |
| Smallest | 20 | 10 | 0 | 0 | 9 | 13 | 108 | 8 | 4 | – |
| MSE-PP | 0.022 | 0.020 | 0.021 | 0.010 | 0.019 | 0.009 | 0.010 | 0.022 | 0.009 | – |
| Flipped | 0.078 | 0.073 | 0.075 | 0.064 | 0.090 | 0.063 | 0.055 | 0.076 | 0.056 | 0.046 |

*Note*: Mean is the simple unweighted average. RE, UWLS, and HS denote random effects, unrestricted weighted least squares, and Hunter and Schmidt approach, respectively. RE and UWLS are numbered by the standard error used; 1 and 2 denote the standard error from Eqs. 2 and 3, respectively. 'z' refers to Fisher's *z* transformations. PP is the PET-PEESE estimate. 'Smallest' displays the frequency that a given estimator is the smallest weighted average, and MSE-PP is the average squared differences between each weighted average and PP. Flipped are the corresponding means for the subset of 50 MAs where the majority of the estimates were negative. 'Corrected' denotes correction for publication bias.

Consistent with past simulations, RE and UWLS are larger when $S_2$, Eq. 3, is used rather than those calculated using $S_1$, Eq. 2. Although the average difference of RE1 and RE2 is quite small, it is nonetheless statistically significant (paired $t$ = 8.76; p<.0001). Also, as expected, the RE estimators are larger than the UWLS family of estimators: UWLS1, UWLS+3, and UWLSz but not as large as the simple unweighted average, the 'Mean'. With PSB, the magnitude of RE has been generally shown to be between UWLS and the unadjusted mean, and RE moves closer to the mean with higher heterogeneity.[27-29] The choice of PCC variance makes a larger difference to the estimated mean effect as calculated by UWLS (0.03) because a few large and highly precise estimated PCCs can carry considerable weight when using $S_2$.[vi] As shown by Stanley and Doucouliagos[10] and corroborated here, there is little reason for the continued use of the conventional PCC standard error formula, $S_2$, in the meta-analysis of PCCs or correlations.[12]

Without $S_2$, the means range from 0.096 to 0.131, while the median varies from 0.056 to 0.088. Note that all medians are smaller than 0.1 and the reduction of the medians relative to their corresponding means is greater than the differences seen across these alternative approaches to the meta-analysis of PCCs. Similar to the ICT illustration, above, RE's and REz's mean and median are both larger than the respective values of HS and the family of UWLS estimators. Again, HS is the smallest, on average, and it is the smallest most frequently (Table 1). Interestingly, the highest correlation (0.997) among these methods is between HS and UWLSz. Although Fisher's $z$ version of random effects has often been suggested as the preferred method,[48] it has the largest mean and median, slightly greater than RE. Therefore, random-effects using Fisher's $z$ is the least conservative and thereby consistent with being the most biased estimate of the mean among these alternative approaches to the meta-analysis of economic PCCs.

One may prudently question whether smaller is truly less biased. Of course, random sampling errors and/or the random influence of one or a few extreme estimates can cause some of these individual estimates across 172 meta-analyses to be smaller (or larger) than the true population mean effect. However, any notable systematic underestimation of the 'true' mean effect by these weighted averages is rather unlikely when they are averaged across 172 meta-analyses and where there is no known downward bias.[11,12] Furthermore, there are clear indications of publication selection bias (PSB) in this survey that, on average, exaggerate the reported magnitudes of mean effects. Nonetheless, we cannot logically infer less bias from smaller average estimates as that would commit the error of 'affirming the consequent.' Rather, these smaller average estimates exhibit a pattern consistent with the 'upward bias hypothesis' and with past simulations where we know that smaller is, in fact, less biased.

To confirm the likely presence of PSB in our survey data, we first ran an aggregate fixed-effect-panel Egger regression with cluster-robust SEs across 172 MAs and 63,671 PCCs. The meta-regression coefficient on SE, S1 from Eq. (2), is 0.850; 95% CI = (0.442, 1.26). This panel meta-regression model allows every area of research to have a different mean effect; however, the PSB effect (estimated to be 0.850 standard errors) is assumed constant across MAs. Although rather small,[26] the PSB effect is clearly positive and statistically significant ($t = 4.08$; $p < 0.001$). In contrast, the proportion of statistically significant test (PSST) allows both the mean effect and the intensity of PSB to vary freely across any number of MAs and presumes no specific mechanism of selection or preference for statistical significance and, thereby, cannot be interpreted as a small-study effect.[47] Also, PSST and its constituent elements (excess statistical significance, ESS, and the expected number of statistically significant reported effects in the absence of any PSB, Esig) are easily, meaningfully, and naturally aggregated across meta-analyses in meta-research analysis.[14,15,47] PSST is generally a more powerful test of PSB as we see here, where PSST $z = 24.2$ ($p << 0.001$). Thus, there is clear evidence for selection for statistical significance in some studies among these 172 meta-analyses. When we look at 172 individual Egger tests for PSB, 47.1% are statistically significant in a positive direction ($\alpha = 0.05$) while 39.5% of PSSTs are statistically significant. Thus, this survey contains clear evidence that is consistent with notable selection for statistical significance of the type that leads to upward biases. Note that the smallest average estimates (HS and the UWLS family of estimators) are also the closest (i.e., smallest MSE) to the PET-PEESE corrected estimates across these 172 meta-analyses (Table 1), which further corroborates an association of the relative magnitudes of these estimators' averages with likely bias. In this context, smaller average estimates are likely to be less biased.

The meta-research findings of these PCC meta-analysis methods, as reported in Table 1 and seen in practice, are fully consistent with relative magnitudes of these estimators seen in previous simulations, including the expectation that HS will be both the smallest and the least biased.[10-12] Although not all of these estimators were simulated, those that were have the same ordering from smallest to largest when there is known publication selection bias. For the smallest population mean (0.11) simulated, the average biases across the 50% and no PSB heterogeneity conditions are: 0.006 for PP, 0.031 (HS), 0.038 ($UWLS_{+3}$), and 0.042 for REz.[12] The relative magnitudes found among our survey's average estimates relative to each other and to PET-PEESE (PP) exactly match what was seen in simulations. However, as expected, the precise magnitudes of these differences cannot be expected to be the same because the observed conditions do not exactly match the simulations. For example, from the aggregate estimates of ESS (excess statistical significance), we estimate that the incidence of PSB is $Psss = \frac{ESS}{1-Esig} = \frac{0.04455}{1-0.3743} = .071$.[15] Though present, this is a notably lower incidence of publication bias than the average of the 50% and no PSB simulation conditions. Thus, we would expect to see smaller apparent PSB biases in this survey than found in past simulations. Because 0.009 is the largest reduction in UWLS seen among either RE's or UWLS's small-sample bias corrections, most of the small differences seen across these alternative MA methods are therefore likely to be driven by PSB.

The strong performance of HS is also consistent with growing evidence that reported precision in observational research can be endogenous to researchers' modeling choices.[41] Recent research focused on the small-sample biases of PCC meta-analyses and showed that they could be driven to negligibility through alternative approaches, REz and $UWLS_{+3}$. Here, we find that these corrections for PCC's make no noticeable reduction to these small-sample biases—compare UWLS to $UWLS_{+3}$, or UWLSz, and RE to REz. Thus, the small differences among methods are likely due to the reduction of PSB as they are also consistent with other simulation studies that compared the performances of RE and UWLS.[24,25,27-32] Also, see the "living synthetic benchmarks" simulation comparisons compiled from over 1,600 previously published simulation designs.[32]

When we focus only on those MAs where the majority of the PCCs < 0 (Flipped, Table 1), all averages are consistently 0.04 to 0.05 smaller. This should not be surprising as these MAs are more likely to be drawn from populations with small or virtually zero mean effects. Nonetheless, we again find the same ordering from smallest to largest among these MA estimators as seen before (Table 1) and in simulations where there is known publication selection bias. The preservation and stability of this complex ordering of estimators' averages is further consistent with the explanation that the observed differences seen in this survey are due to their known differential responses to PSB and thereby consistent with the upward bias hypothesis.

Although often small, all of the differences among these meta-analysis methods are statistically significant— REz v RE (0.005; paired $t$ = 4.87 p < .0001), RE v UWLS (0.015; $t$ = 3.56; p < .001), UWLS v UWLS$_{+3}$ (0.001; $t$ = 6.67; p < .0001), UWLS v UWLSz (0.009; $t$ = 4.71; p < .0001), RE v HS (0.029; $t$ = 7.09; p < .0001), UWLSz v HS (0.005; $t$ = 4.71; p < .0001), and UWLS$_{+3}$ v HS (0.013; $t$ = 5.48; p < .0001). However, are these small numerical differences in correlations practically or scientifically notable?

The smallest effect size of interest is context specific. Nonetheless, a Cohen's $d$ of 0.2 is commonly used when lacking specific theoretical or policy guidance to the contrary. Because we are evaluating 172 very different areas of research without a common cost-benefit profile, we have no choice but to default to widespread practice, which is Cohen's guidelines for a 'small' (or nontrivial) correlation of 0.1 (equivalent to Cohen's $d$ < 0.2).[46] Partial correlations less than 0.1 contribute less than 1%, marginally, to the explanation of the phenomenon in question. With a few exceptions where there are large-scale and low-cost applications, PCCs smaller than 0.1 would be widely regarded as trivial. In a field as noisy as observational economics research, accounting for less than 1% of the relevant variation is trivial. By this guideline, the majority of economics research investigates empirically trivial phenomena, at least as represented by these 172 meta-analyses. Relative to RE, 55% of these areas of research are scientifically trivial, 66% for UWLS and UWLS$_{+3}$, and 67% for UWLSz and HS.

How often do these corrections make a notable difference in how we might interpret the average effect in an area of economics research? That is, how often would a nontrivial average effect as calculated by RE, the conventional approach, become scientifically trivial when these small-sample bias reductions are applied? 32 or 33% of small RE estimates ($0.1 < r_p < 0.3$) are reduced to triviality when replaced by UWLS, UWLS$_{+3}$, UWLSz, or HS, while the reverse happens once, < 2%. Thus, UWLS and HS can make a notable difference at the margins in practice.

An important limitation is that all of these alternative MA estimators are estimates of the *unconditional* mean effect. "(A)ny simple overall meta-analysis needs to be interpreted with caution if there is publication selection bias and/or heterogeneity in the reported estimates".[7(p713)] Nonetheless, high heterogeneity is typical of economics and the social sciences. According to our PCC survey, the average $I^2$ is 86% and $\hat{\tau}^2$ is 0.026. All but one MA has statistically significant heterogeneity. With heterogeneity this high, one might question whether any single mean estimate can be representative in the majority of these areas of research. Because notable heterogeneity is widely recognized and generally assumed by economists, almost all economic meta-analyses code dozens of research dimensions to accommodate heterogeneity through multiple meta-regression analysis (MRA), thereby providing conditional estimates of mean effect. Although any simple overall average may not be relevant to many economic meta-analyses, they remain relevant to medicine and psychology, where mainly unconditional meta-analysis averages are reported. Even for economics, biases found among the unconditional mean estimates are likely to propagate into the MRA, thereby distorting the estimated effects of moderators. The central point to this survey is to understand whether the small-sample biases uncovered in recent studies are likely to affect the interpretation of meta-analysis findings in practice, and we find that they generally do not.

Another potential caveat is that the typical PCC in these 172 meta-analyses is quite small, median average PCC = 0.073, and small-sample biases are larger with larger PCCs.[10,11] However, even for the larger PCC meta-analyses, we find that the small-sample bias is negligible. Only one meta-analysis has a difference between UWLS and $UWLS_{+3}$ > 0.01 (0.011). Nonetheless, it would be prudent for meta-analyses of large PCCs to routinely use one of these corrections for small-sample bias as a robustness check.

UWLS, UWLS+3, and UWLSz can be directly incorporated into MRA without any further alterations. However, there is no known MRA extension for HS. Developing such an extension is a promising direction for future research. Thus, UWLS is likely the most practical PCC approach for economics applications, with estimates consistent with smaller upward bias, and HS or UWLSz is likely to be the best estimate of mean correlation in medicine and psychology. As a simple rule of thumb, we recommend that meta-analysts of PCCs in any discipline report UWLS and focus on HS or UWLSz when only an unconditional mean is needed.

## 7. CONCLUSION

This study surveys the findings from alternative meta-analysis methods across 172 meta-analyses of partial correlation coefficients, PCCs. The primary purpose of this survey is to assess whether recently identified biases of PCC meta-analyses and the associated methods offered to reduce these biases would likely make a noticeable difference to the practice of meta-analysis.[9-11]

Van Aert and Goos[9] argued that meta-analyses using standard errors calculated from Eq. (2), $S_1$, would be biased relative to using the 'correct' standard errors, $S_2$, Eq. (3). Stanley and Doucouliagos[10] demonstrated that the exact opposite was true; that is, conventional meta-analysis methods (RE, FE, and UWLS) using $S_1$ have consistently smaller small-sample biases than those calculated from $S_2$. This survey confirms the advantage of using $S_1$ rather than $S_2$ as the PCC's SE because the averages for both RE and UWLS across 172 PCC meta-analyses increase when $S_2$ is used in place of $S_1$. Recall that the way these meta-analyses are aligned (inverse associations made positive) implies that both small-sample bias and publication selection bias (PSB) if present will exaggerate the magnitude of reported effects. The estimators with smaller average effect sizes (those using $S_1$) are known to be less biased in simulations and thereby are consistent with smaller biases especially as there is ample evidence of PSB in our survey data. Because these empirical findings are consistent with theory and simulations, we see little reason to continue using the conventional standard error, $S_2$, in the meta-analyses of PCCs.

To correct the small-sample biases identified in recent studies, Stanley, Doucouliagos and Havranek[11] suggested a simple correction for degrees of freedom, $UWLS_{+3}$, and the well-known Fisher's *z* because its variance does not depend on the PCC estimate. However, these corrections, REz, UWLSz, and $UWLS_{+3}$, do not notably differ from their corresponding unadjusted meta-analyses, RE and UWLS, and thereby make little scientific or practical difference to how PCC meta-analyses would be interpreted or understood in practice. As discussed above, the sample sizes and complexity of these 172 PCC meta-analyses wash out any notable small-sample bias. What little remains is likely publication selection bias (PSB). This is consistent with the fact that UWLS is smaller and less biased than the corresponding random effects estimator in simulations when there is known PSB. In particular, UWLS's median is 30% smaller than RE's, consistent with the relative performance of these estimators found in numerous simulation studies when there is PSB.[24,25,27-30,32] Also, notable PSB is confirmed by the proportion of statistically significant test (PSST)[47] and the Egger meta-regression.

Stanley, Doucouliagos and Havranek[12] also investigated the Hunter and Schmidt (HS) approach to the meta-analysis of bivariate correlations. In this survey of PCC meta-analyses, HS is the smallest and thereby consistent with the upward bias hypothesis' association of smaller with less biased and with Hunter and Schmidt's[13] argument that their meta-analysis approach is preferable to RE using Fisher's $z$. Although HS is typically a little smaller than the UWLS family of estimators, this difference will rarely be meaningful in practice. Besides, the median of HS and UWLSz are the same. As shown in previous simulations, meta-research, and consistent with this survey, HS and the UWLS family generally have superior statistical properties relative to RE.[10-12,24,25,27-30,32,49] However, extreme estimates can differentially affect these methods in specific areas of research. Thus, the more conservative (i.e., smaller and less statistically significant) estimate should be emphasized in summary discussions if in doubt.

The only meaningful difference among any of these estimators is at the margins when compared to RE. When 0.1 is regarded as the smallest effect size of interest, as discussed above, approximately one-third of the 'small' average effects identified by RE become 'negligible' by UWLS, $UWLS_{+3}$, UWLSz, and HS.

In summary, this survey supports the use of $S_1$ over $S_2$ and UWLS and HS over RE in the meta-analysis of partial correlation coefficients. Otherwise, these corrections for the small-sample bias of the meta-analysis of partial correlation coefficients matter little in practice, at least as seen in our large sample of economic meta-analyses. However, these differences may still be of consequence for some areas of experimental research where the sample sizes are routinely small. We encourage researchers in health sciences, psychology, and environmental studies who meta-analyze PCCs to compare these estimators using data from their own fields.

## REFERENCES

1. Ioannidis JPA. The mass production of redundant, misleading, and conflicted systematic reviews and meta-analyses. *Milbank Q.* 2016;94:485-514.

2. Owens DK, Lohr KN, Atkins D, et al. AHRQ series paper 5: grading the strength of a body of evidence when comparing medical interventions. *J Clin Epidemiol.* 2010;63(5):513-523.

3. Stanley TD. Wheat from chaff: meta-analysis as quantitative literature review. *J Econ Perspect.* 2001;15:131-150.

4. Borenstein M, Hedges LV, Higgins JPT, Rothstein HR. *Introduction to Meta-Analysis.* John Wiley and Sons; 2009.

5. Stanley TD, Doucouliagos H. *Meta-Regression Analysis in Economics and Business.* Routledge; 2012.

6. Havranek T, Irsova Z, Zeynalova O. Tuition fees and university enrolment: a meta-regression analysis. *Oxf Bull Econ Stat.* 2018;80:1145-1184.

7. Stanley TD, Doucouliagos H, Steel P. Does ICT generate economic growth? A meta-regression analysis. *J Econ Surv.* 2018;32:705-726.

8. Geyer-Klingeberg J, Hang M, Rathgeber AW. What drives financial hedging? A meta-regression analysis of corporate hedging determinants. *Int Rev Financ Anal.* 2019;61:203-221.

9. van Aert RCM, Goos C. A critical reflection on computing the sampling variance of the partial correlation coefficient. *Res Synth Methods.* 2023;14:520-525.

10. Stanley TD, Doucouliagos H. Correct standard errors can bias meta-analysis. *Res Synth Methods.* 2023;14:515-519.

11. Stanley TD, Doucouliagos H, Havranek T. Meta-analyses of partial correlations are biased: detection and solutions. *Res Synth Methods.* 2024;15:313-325.

12. Stanley TD, Doucouliagos H, Havranek T. Reducing the biases of the conventional meta-analysis of correlations. *Res Synth Methods.* 2025;16:42-59.

13. Hunter JE, Schmidt FL. *Methods of Meta-Analysis: Correcting Error and Bias in Research Findings.* Sage; 1990.

14. Askarov Z, Doucouliagos A, Doucouliagos H, Stanley TD. The significance of data-sharing policy. *J Eur Econ Assoc.* 2023; *21:*1191–1226.

15. Askarov Z, Doucouliagos A, Doucouliagos H, Stanley TD. Selective and (mis)leading economics journals: meta-research evidence. *J Econ Surv.* 2024;38:1567-1592.

16. Havránek T, Stanley TD, Doucouliagos H, et al. Reporting guidelines for meta-analysis in economics. *J Econ Surv.* 2020;34:469-475.

17. Irsova Z, Doucouliagos H, Havranek T, Stanley TD. Meta-analysis of social science research: a practitioner's guide. *J Econ Surv.* 2024;38:1547-1566.

18. Gustafson RL. Partial correlations in regression computations. *J Am Stat Assoc.* 1961;56:363-367.

19. Olkin I, Siotani M. Asymptotic distribution of functions of a correlation matrix. In: Ikeda S, ed. *Essays in Probability and Statistics.* Shinko Tsusho; 1976:235-251.

20. Brodeur A, Lé M, Sangnier M, Zylberberg Y. Star Wars: the empirics strike back. *Am Econ J Appl Econ.* 2016;8:1-32.
21. Bruns SB, Deressa TK, Stanley TD, Doucouliagos C, Ioannidis JPA. Estimating the extent of selective reporting: an application to economics. *Res Synth Methods.* 2024;15:590-602.
22. Wooldridge JM. *Introductory Econometrics.* 2nd ed. Thomson/South-Western; 2003.
23. Ioannidis JPA, Stanley TD, Doucouliagos C. The power of bias in economics research. *Econ J.* 2017;127:F236-F265.
24. Poole C, Greenland S. Random-effects meta-analyses are not always conservative. *Am J Epidemiol.* 1999;150:469-475.
25. Henmi M, Copas JB. Confidence intervals for random effects meta-analysis and robustness to publication bias. *Stat Med.* 2010;29:2969-2983.
26. Doucouliagos C, Stanley TD. Theory competition and selectivity: are all economic facts greatly exaggerated? *J Econ Surv.* 2013;27:316-339.
27. Stanley TD, Doucouliagos H. Neither fixed nor random: weighted least squares meta-analysis. *Stat Med.* 2015;34:2116-2127.
28. Stanley TD, Doucouliagos H. Neither fixed nor random: weighted least squares meta-regression analysis. *Res Synth Methods.* 2017;8:19-42.
29. Stanley TD, Doucouliagos H, Ioannidis JPA. Finding the power to reduce publication bias. *Stat Med.* 2017;36:1580-1598.
30. Bom PRD, Rachinger H. A kinked meta-regression model for publication bias correction. *Res Synth Methods.* 2019;10:497-514.
31. Bartoš F, Maier M, Wagenmakers EJ, et al. Footprint of publication selection bias on meta-analyses in medicine, environmental sciences, psychology, and economics. *Res Synth Methods.* 2024;15:500-511.
32. Bartoš F, Pawel S, Siepe BS. Living synthetic benchmarks: a neutral and cumulative framework for simulation studies. Preprint. Posted online 2025. doi:10.48550/arXiv.2510.19489.
33. Hartung J, Knapp G. On tests of the overall treatment effect in meta-analysis with normally distributed responses. *Stat Med.* 2001;20:1771-1782.
34. Sidik K, Jonkman JN. A comparison of heterogeneity variance estimators in combining results of studies. *Stat Med.* 2007;26:1964-1981.

35. van Aert RCM, Jackson D. A new justification of the Hartung-Knapp method for random-effects meta-analysis based on weighted least squares regression. *Res Synth Methods.* 2019;10:515-527.
36. Greene WH. *Econometric Analysis.* Macmillan; 1990.
37. Davidson R, MacKinnon JG. *Econometric Theory and Methods.* Oxford University Press; 2004.
38. Stanley TD, Ioannidis JPA, Maier M, Doucouliagos H, Otte WM, Bartoš F. Why the unrestricted weighted least squares should be routinely reported in medical meta-analyses. Preprint. Posted online 2026. https://osf.io/vpuqj/files/4rkqg
39. Fisher RA. The distribution of the partial correlation coefficient. *Metron.* 1924;3:329-332.
40. Field AP. Meta-analysis of correlation coefficients: a Monte Carlo comparison of fixed- and random-effects methods. *Psychol Methods.* 2001;6:161-180.
41. Irsova Z, Bom PRD, Havranek T, Rachinger H. Spurious precision in meta-analysis of observational research. *Nat Commun.* 2025;16:4324.
42. Berinsky AJ, Druckman JN, Yamamoto T. Publication biases in replication studies. *Polit Anal.* 2021;29(3):370-384.
43. Hume D. *A Treatise of Human Nature.* Selby-Bigge LA, Nidditch PH, eds. Oxford University Press; 1978. Originally published 1739-1740.
44. Popper K. *The Logic of Scientific Discovery.* 1959.
45. Solow RM. We'd better watch out. *New York Times Book Review.* July 12, 1987:36.
46. Cohen J. *Statistical Power Analysis for the Behavioral Sciences.* 2nd ed. Academic Press; 1988.
47. Stanley TD, Doucouliagos H, Ioannidis JPA, Carter E. Detecting publication selection bias through excess statistical significance. *Res Synth Methods.* 2021;12:776-795.
48. van Aert RCM. Meta-analyzing partial correlation coefficients using Fisher's z transformation. *Res Synth Methods.* 2023;14:768-773.
49. Stanley TD, Ioannidis JPA, Maier M, Doucouliagos H, Otte WM, Bartoš F. Unrestricted weighted least squares represent medical research better than random effects in 67,308 Cochrane meta-analyses. *J Clin Epidemiol.* 2023;157:53-58.

**Competing Interests Statement**

None

**Funding statement**

Grantová Agentura České Republiky: 23-05227M & 24-11583S

**Data Availability Statement:**

The data and code used in this survey are available at: https://github.com/PetrCala/pcc-survey

---

[i] 'Fixed effect' in a meta-analysis context has an entirely different meaning than 'fixed effects' in panel data models. In panel meta-regression models, fixed effects are the general case while random effects are applicable only when study effects are confined to be normally distributed and the meta-regression coefficients are constant across studies.[5,22] Fixed-effects meta-meta-models can offer stronger quasi-experimental designs for meta-research.[14]

[ii] We thank an anonymous reviewer for suggesting that we report the UWLS version of using Fisher's $z$ transformation. It proves to be, arguably, the best approach in practice—see below.

[iii] It would also be logical to replace sample sizes, $n_i$, with PCCs degrees of freedom, potentially improving HS's mean squared error. However, we also suspect that such nuances will not matter in practical applications which have much larger biases (PSB) and inefficiencies. In the 172 meta-analyses that we investigate below, we use degrees of freedom as not all authors reported sample size. Fortunately, degrees of freedom can always be calculated from Eq. (1) when we know $r_p$ and its SE, both of which are required in meta-analyses.

[iv] It is not clear that the unconditional mean in this application carries much inferential value as heterogeneity is quite high. Because these estimators are averaged across PCCs employing very different measures of economic growth (per capita GDP, GDP growth rates, and productivity), it is unclear whether any mean is a meaningful representation of this area of research. The meta-analysis from which this illustration is taken agrees. "(A)ny simple overall meta-analysis needs to be interpreted with caution if there is publication selection bias and/or heterogeneity in the reported estimates".[7(p713)] Almost all economic meta-analyses of PCCs, including this one, are primarily meta-regression analyses (MRA) that explicitly control for observed differences of measures and methods, among other factors, and rely on the MRA, conditional, findings for their conclusions. Here, we are merely illustrating the typical pattern found in simulations among these alternative MA estimators and across 172 MAs of PCCs, as discussed below. We make no inference about the 'true' unconditional average PCC of ICT.

[v] We thank an anonymous reviewer for suggesting that we calculate the 'smallest' and the closeness to PET-PEESE.

[vi] When one or few estimates notably change RE or UWLS, they are potentially influence (and leverage) points and should be further investigated. Experience suggests that the specific estimate that causes notable changes to the magnitude of the average effect is often a coding error, typo, or a conceptually different but related estimate. Because the purpose of this empirical study is to survey how different MA methods affect PCC meta-analyses in practice, we did not attempt to further remove outliers or influence points.